\documentclass[12pt]{article}

\usepackage{latexsym}
\usepackage{graphicx}
\usepackage{rotating}
\usepackage[normalem]{ulem}
\usepackage{hyperref}
\usepackage[version=4]{mhchem}
\usepackage{amsmath,amssymb,amsfonts}
\usepackage{bm}
\usepackage{color}
\usepackage{soul}

\usepackage{color}

\newcommand{\lyxmathsym}[1]{\ifmmode\begingroup\def\b@ld{bold}
        \text{\ifx\math@version\b@ld\bfseries\fi#1}\endgroup\else#1\fi}

\usepackage{amsfonts}
\usepackage{mathrsfs}
\usepackage[latin9]{inputenc}
\usepackage{color}
\definecolor{note_fontcolor}{rgb}{0.800781, 0.800781, 0.800781}
\usepackage{mathrsfs}
\usepackage{amsmath}
\usepackage{amssymb}
\usepackage{graphicx}
\hypersetup{colorlinks=true,urlcolor=blue,linkcolor=blue}

\usepackage{scicite}

\usepackage{times}

\newenvironment{sciabstract}{%
        \begin{quote} \bf}
        {\end{quote}}

\title{Evening out the spin and charge parity to increase T$_c$ in unconventional superconductor \ce{Sr2RuO4}}
\author
{
        Swagata Acharya$^{1\ast}$, Dimitar Pashov$^{1}$, C\'edric Weber$^{1}$, \\
        Hyowon Park$^{2,3}$, Lorenzo Sponza$^{4}$, Mark van Schilfgaarde$^{1}$ \\
        \\
        \normalsize{$^{1}$King's College London, The Strand, WC2R 2LS London, UK,}\\
        \normalsize{$^{2}$Department of Physics, University of Illinois at Chicago, Chicago, Illinois 60607, USA},\\
        \normalsize{$^{3}$Materials Science Division, Argonne National Laboratory, Argonne, Illinois, 60439, USA},\\
        \normalsize{$^{4}$LEM UMR 104, ONERA-CNRS, F-92322, Ch\^{a}tillon, France }\\
        \\
        \normalsize{$^\ast$E-mail:  swagata.acharya@kcl.ac.uk.}
}

\date{}

\begin{document}

\baselineskip24pt


\maketitle

\begin{sciabstract}
Unconventional superconductivity in Sr$_{2}$RuO$_{4}$ has been intensively studied for decades. The origin and nature of
the pairing continues to be widely debated, in particular, the possibility of a triplet origin of Cooper pairs.  However, complexity of
Sr$_{2}$RuO$_{4}$ with multiple low-energy scales, involving subtle interplay among spin, charge and orbital degrees of
freedom, calls for advanced theoretical approaches which treat on equal footing all electronic effects. Here we develop
a novel approach, a detailed \emph{ab initio} theory, coupling quasiparticle self-consistent \emph{GW} approximation
with dynamical mean field theory (DMFT), including both local and non-local correlations. We report that the
superconducting instability has multiple triplet and singlet components. In the unstrained case
the triplet eigenvalues are larger than the singlets. Under uniaxial strain, the triplet
eigenvalues drop rapidly and the singlet components increase. This is concomitant with our observation of spin and
charge fluctuations shifting closer to wave-vectors favoring singlet pairing in the Brillouin zone. We identify a complex
mechanism where charge fluctuations and spin fluctuations co-operate in the even-parity channel under strain leading to increment in $T_c$, thus proposing a novel
mechanism for pushing the frontier of $T_c$ in unconventional `triplet' superconductors.
\end{sciabstract}



Superconductivity~\cite{onnes} is a quantum phenomenon where electrons participate in
dissipation-less charge transport. While electrons repel each other via the Coulomb force, quantum theory provides additional interactions, that in special circumstances at low temperature, can overcome the repulsion to bind electrons in 
Cooper pairs~\cite{cooper}. Within the paradigm of unconventional superconductivity the primary glue for the Cooper pairing can originate from collective bosonic excitations other than phonons.  It usually involves spin fluctuations, but a general understanding of its origin is lacking.  Here we focus on \ce{Sr2RuO4} (SRO), an unconventional superconductor which is highly sensitive to disorder~\cite{disorder98}.  SRO is
of great interest because there are indications that superconductivity has spin-triplet symmetry, which raises the possibility that it can sustain Majorana states conducive for topological quantum computing~\cite{nayak}.


SRO single crystals were first shown to exhibit superconductivity below 1.5 K in 1994~\cite{maeno94}; yet the origin for
pairing is still debated~\cite{odder}.  The superconducting transition temperature, $T_{c}$, has been observed to
increase to 3\,K in eutectic crystals of SRO, in the vicinity of Ru
inclusions~\cite{eutectic98,eutectic09,eutectic-09}. While enhancement of $T_{c}$ was traditionally associated with a
reduced volume fraction, a recent series of experiments \cite{hicks,steppke} on bulk single crystals of SRO subject to
uniaxial strain, show an increase to 3.4 K for compressive strain in the [100] direction, which we denote as
$\epsilon_x$.  These apparently dissimilar studies hint towards a more common underlying mechanism for enhancement of
$T_{c}$, since Ru inclusions induce local stresses which include uniaxial strain. In the tensile experiments $T_{c}$ can
be controlled by varying $\epsilon_x$.  It reaches a maximum value at $\epsilon_x$=0.6\%~\cite{steppke}, beyond which it
falls rapidly.  In what follows we will denote $\epsilon_x$=0.6\% as $\epsilon^*_x$.

These observations challenge the established belief that SRO is a spin-triplet (odd-parity) superconductor.  Under
strain, the tetragonal symmetry of the compound is lost, and it is no longer possible to find a triplet order parameter
from two degenerate components, such as the usual p$_{x}$+ip$_{y}$ or d$_{xz}$+id$_{yz}$. This raises the
possibility for alternative mechanisms that could be responsible for pairing under strain.

The effect of strain on the Fermi surface has been studied with density functional theory (DFT)~\cite{steppke}, and
complementary mimimal model Hamiltonian approaches~\cite{imai,liu}, which identified a change in Fermi surface topology.
In particular, a Van Hove singularity~\cite{steppke} approaches the Fermi level, with a concomitant increase in charge
carriers, which has been suggested as a possible mechanism for the increment in $T_c$~\cite{barber} under strain. Such a
picture identifies an important property resulting from strain, but it is not sufficient to explain the enhancement of
$T_{c}$.  In particular, the multi-orbital nature of the spin and charge fluctuations and many-body correlations are
shown to be important in SRO~\cite{mravlje11,damascelli14,swagata17,lechermann18}.  Novel electron correlations
originating from competition between non-local Coulomb repulsion and the large Hund's
coupling~\cite{baskaran,mravlje11,swagata18} are also significant.


It is a formidable challenge to adequately describe the single- and two-particle responses needed for insights into the
origin and nature of superconductivity in SRO.  As we show here, an accurate theoretical formulation, that
includes both local and non-local correlation effects in space, momentum and time and for all relevant degrees of
freedom, is essential.  Recently, a significant advance has been achieved by combining the quasi-particle self
consistent \emph{GW} (QS\emph{GW}) approximation, with dynamical mean field theory (DMFT)~\cite{prx18,prbr17}.  Merging
of these two state-of-the-art methods captures the effect of both strong local dynamic spin fluctuations (captured well
in DMFT), and non-local dynamic correlation~\cite{tomc} effects (captured by QS\emph{GW}), relying on neither model
hamiltonians, nor on DFT, and avoiding the concomitant limitations they each carry.  Also, nonlocal spin, charge and
pairing susceptibilities can also be obtained from vertices computed from the local two-particle Green's sampled by DMFT
and bubble diagrams, via the solutions of Bethe Salpeter equations in respective channels. The full numerical implementation is discussed in Pashov et. al.~\cite{questaal-paper} and codes are available on the open source electron structure suite Questaal~\cite{questaal_web}.

Here we apply this new methodology to SRO, studying the pristine compound and also the effect of strain.  Through the
vertices and susceptibilities we can identify what drives superconductivity, and also what causes the non-monotonic
dependence of $T_{c}$ on strain.  The pairing instability has multiple singlet and triplet components; nodal structure in singlet channel and nodeless odd-frequency structure in triplet channel. We find that the pairing is favored by even parity couplings in both spin and charge channels as $\epsilon_x$ approaches $\epsilon_x^*$ from below, while for $\epsilon_x > \epsilon_x^*$ incoherent spin fluctuations suppress the superconducting order.  Our observations are in remarkable agreement with recent neutron scattering
experiments~\cite{maeno18}.



\begin{figure}
        \begin{center} \includegraphics[width=1.00\columnwidth]{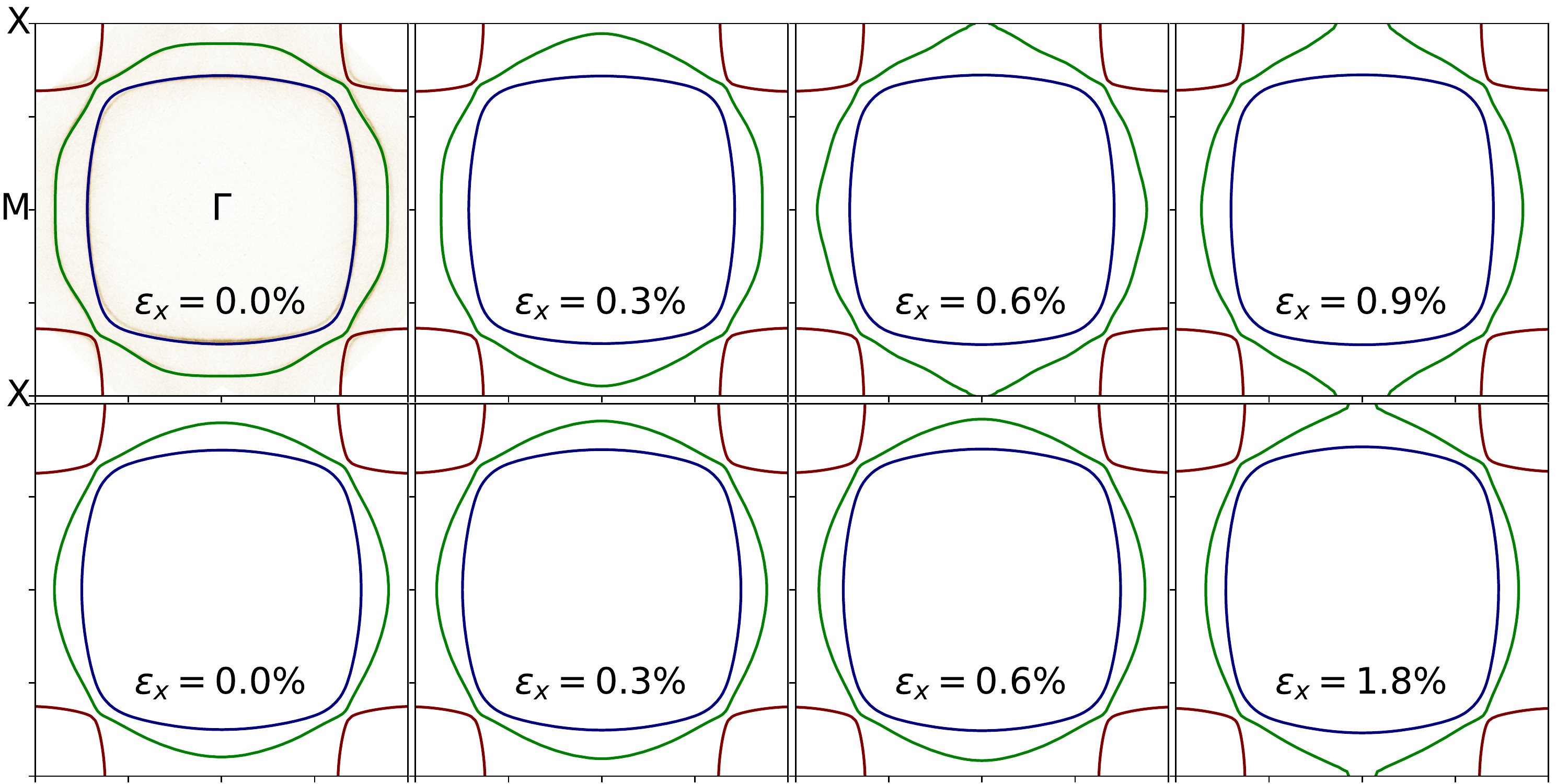}
          \caption{\textbf{Evolution of Fermi surface topology under strain:} Top row shows the
            QS\emph{GW} Fermi surfaces in the basal plane, for a [100] compressive strain with $\epsilon_x$=(0\%,
            0.3\%, 0.6\%, 0.9\%).  Spin orbit coupling is included (its omission makes a modest change to the Fermi
            surfaces).  In the first panel (top left) high-resolution ARPES data~\cite{tamai2018} (figure is replotted using the raw ARPES data) for the Fermi surfaces are shown (figure reproduced with due permission) in the background of our QS\emph{GW} theoretical data. For a higher resolution comparison please see the SM. States derive almost exclusively of Ru $t_{2g}$ orbitals $xy$, $xz$, $yz$; the orbital character
            of each pocket changes moving around the contour.  $xy$ character is present on the entire Fermi surface: it
            resides on the blue pocket on the $\Gamma$-X line, and on the green on the $\Gamma$-M line.  Under strain,
            the four M points lose the 4-fold rotational symmetry, and at $\epsilon_x=\epsilon_x^*$ the topology of the
            green band changes.  Bottom row shows corresponding results for DFT.  In DFT the transition also occurs, but
            near $\epsilon_x$=1.8\% (bottom right panel), instead of $\epsilon_x^*$. }    
        \label{fig:fermi}
        \end{center}
\end{figure}

\emph{Evolution of Fermi surface topology under strain:} Fermi surfaces in the basal plane are shown in
Fig.~\ref{fig:fermi}.  The critical change in topology on the line connecting (0,0) and (0,$\pi$) (points $\Gamma$ and
M) occurs at $\epsilon_x$=0.6\%, in excellent agreement with $\epsilon_x^*$.  This is the strain where one Van-Hove
singularity crosses $E_{F}$ (see SM), as was noted in a prior DFT study; though in DFT it does so at a much larger
$\epsilon_x$ (see Fig.~\ref{fig:fermi}).  The Fermi surface generated by QS\emph{GW} closely matches recent
high-resolution bulk Fermi surface observed in quantum oscillation studies~\cite{bergemann} angle-resolved
photo-emission spectroscopy~\cite{tamai2018,dai}; indeed they can only be easily distinguished at at higher resolution
than Fig.~\ref{fig:fermi} (as shown in SM, Fig.~1). That QS\emph{GW} simultaneously yields the topology change close to the
observed $\epsilon_x^*$, and can reproduce fine details of the ARPES Fermi surface, is a reflection on its superior
ability to generate good effective noninteracting Hamiltonians.

\emph{Spin fluctuations: incommensurability and coherence:} 
{Spin ($\chi^{m}$) and charge ($\chi^{d}$) susceptibilities are computed from momentum dependent
  Bethe-Salpeter equations~\ref{eq:BSE_imag} in magnetic (spin) and density (charge) channels.}
\begin{equation}
\chi_{{\alpha_{1},\alpha_{2}\atop \alpha_{3},\alpha_{4}}}^{m(d)}(i\nu,i\nu^{\prime})_{\textbf{q},i\omega}=
[(\chi^{0})_{\textbf{q},i\omega}^{-1}-\Gamma_{loc}^{irr,m(d)}]_{{\alpha_{1},\alpha_{2}\atop \alpha_{3},\alpha_{4}}}^{-1}(i\nu,i\nu^{\prime})_{\textbf{q},i\omega}.
\label{eq:BSE_imag}
\end{equation}
$\chi^{0}$ is the non-local ($k$-dependent ) polarization bubble computed from single-particle QS\emph{GW} Green's
functions dressed by DMFT and $\Gamma$ is the local irreducible two-particle vertex functions computed in
magnetic and density channels.  $\Gamma$ is a function of two fermionic frequencies $\nu$ and $\nu'$ and the bosonic
frequency $\omega$.  The susceptibilities $\chi^{m(d)}(\textbf{q},i\omega)$ are computed by closing
$\chi_{{\alpha_{1},\alpha_{2}\atop \alpha_{1},\alpha_{2}}}^{m(d)}(i\nu,i\nu^{\prime})_{\textbf{q},i\omega}$ with spin or
charge bare vertex $\gamma$ ($\gamma$=1/2 for spin and $\gamma$=1 for charge) and summing over frequencies
($i\nu$,$i\nu^{\prime}$) and orbitals ($\alpha_{1,2}$) (see SM for derivations).
\begin{equation}
\chi^{m(d)}(\textbf{q},i\omega)=
2\mbox{\ensuremath{\gamma}}^{2}\sum_{i\nu,i\nu^{\prime}}\sum_{\alpha_{1}\alpha_{2}}\chi_{{\alpha_{1},\alpha_{2}\atop \alpha_{1},\alpha_{2}}}^{m(d)}(i\nu,i\nu^{\prime})_{\textbf{q},i\omega}.\label{eq:chi}
\end{equation}

\begin{figure}
        \begin{center}
                \includegraphics[width=1.0\columnwidth]{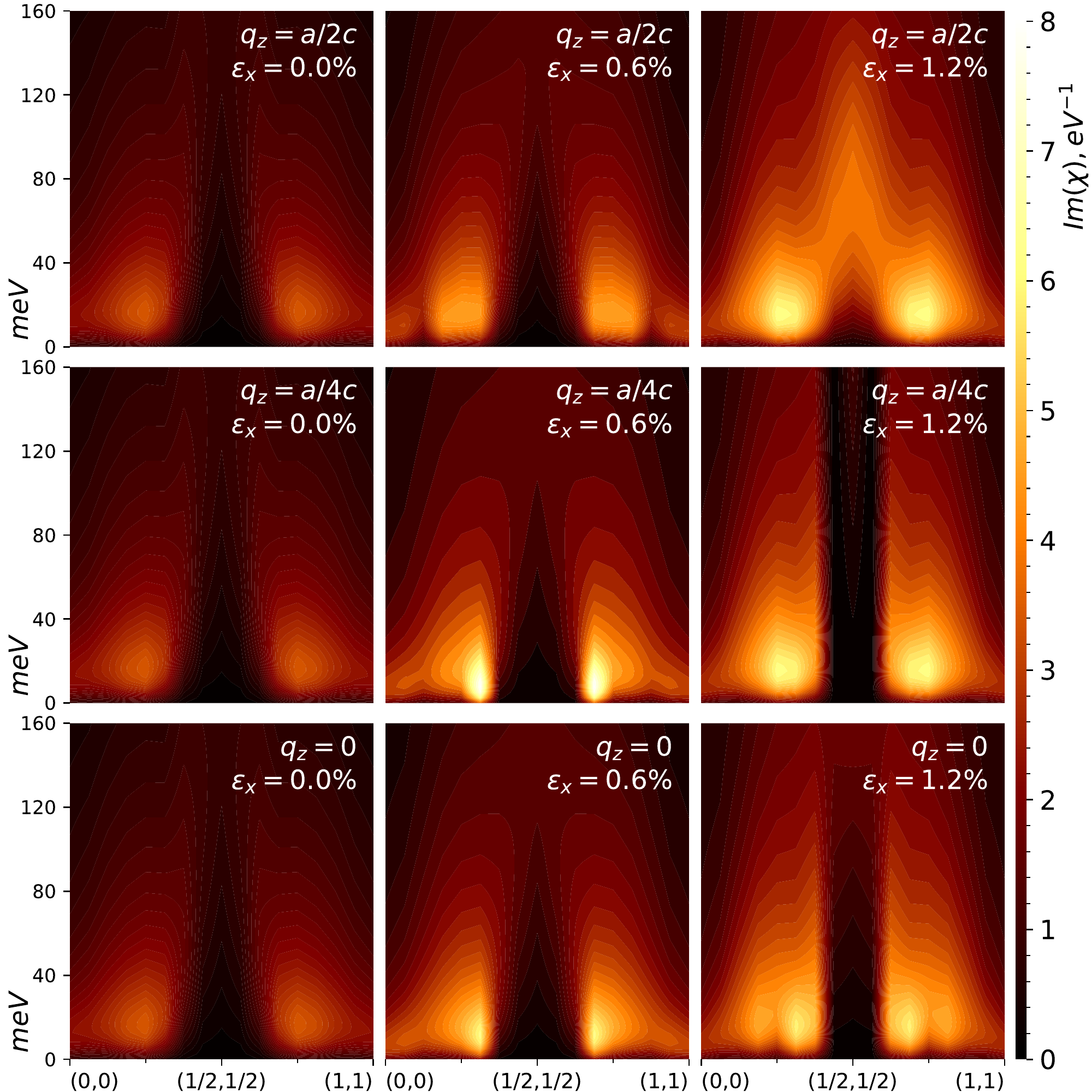}
                \caption{
                                {\bf Spin fluctuations: incommensurability and coherence:} Imaginary part of the dynamic spin susceptibility $\chi^{s}(q,\omega)$ are
                shown in the Cartesian \emph{xy} plane at different values $q_z$, and for different strains $\epsilon_x$.  
                                The unstrained compound shows a spin fluctuation spectrum strongly peaked at $(0.3,0.3,q_z)$ (units
                $2\pi/a$).  At $\epsilon_x$=0, $\chi^{s}$ is nearly independent of $q_{z}$, but it begins to depend on $q_z$
                for $\epsilon_x{>}0$.  With increasing strain fluctuations become more coherent and strongly peaked,
                reaching a zenith at $\epsilon_x{=}\epsilon_x^*$ (0.6\%), where $T_{c}$ is maximum.  For
                $\epsilon_x{>}\epsilon_x^*$, this peak becomes more diffuse; also a secondary incoherent peak emerges at
                $(0.15,0.15,q_z)$, and the quasi anti-ferromagnetic vector $(1/2,1/2,a/2c)$ acquires spectral weight
                around $\omega{=}40$\,meV.
                Note also the spectral weight near the FM vector (0,0,0), and its evolution with $\epsilon_x$.
                }
                \label{fig:spinfluctuations}
        \end{center}
\end{figure}

We focus first on the $\Gamma$-X line of the Brillouin zone, where peaks appear in inelastic neutron scattering
measurements~\cite{braden,sidis,ishida} at the incommensurate vector
$q^\mathrm{IC}{=}(0.3, 0.3, 0)$ (in units $2\pi/a$) which a maximum in frequency near $\omega$=10\,meV.  Using DMFT, we
compute $\chi^s(\mathbf{q},\omega)$ by obtaining the local two-particle vertex in the spin channel and solving the
Bethe-Salpeter equation~\cite{hywon_vertex}, for varying amounts of strain.  Consider first the unstrained case, when
measurements are available. Fig.~\ref{fig:spinfluctuations} shows $\chi^s(q,\omega)$ on the $\Gamma$-X line and in
planes $q_{z}$ = 0, 1/4, and 1/2 (in units of $2\pi/c$). The peak noted above $[q^\mathrm{IC}{=}(0.3, 0.3, 0),\ 
\omega{=}10$\,meV] is nearly independent of $q_z$, and moreover it disperses all the way up to 80 meV.  All of these
findings are in excellent agreement with experimental observations~\cite{Iida}.  We also find significant spin
fluctuations at the ferromagnetic (FM) vector $q{=}(0,0,0)$ (also seen in very recent neutron
measurements\cite{maeno18}) and almost no intensity at the antiferromagnetic nesting vector (1/2,1/2,0).  The FM signal
is important, because of its implications for superconductivity~\cite{maeno18} and whether the pairing is of triplet or
singlet character.  We find that the intensity of $\chi^s(q{=}0)$ is $\sim$1/5 of the dominant IC peak when spin-orbit
(SO) coupling is suppressed.  But SO coupling lifts band degeneracies at high symmetry points, reducing this ratio
slightly to $\sim$1/7.  Thus $\chi^{s}$ seems to be dominated by fluctuations at the IC vector.  Hence, such spin
fluctuation spectra should favour pairing mainly in the spin singlet channel, absent other channels to provide extra
glue for a triplet pairing. However, the continuum of spin excitation prevalent in most part of the Brillouin zone also
can contribute to the glue.

As strain is applied $\mathrm{Im}\chi^s(q,\omega)$ becomes sharper and more coherent, reaching a maximum coherence at
$\epsilon_x{=}\epsilon_x^*$: the peak intensity remains at $q^\mathrm{IC}$ but nearly doubles in strength and shifts to
slightly smaller $\omega$.  Also $\mathrm{Im}\chi^s$ loses its two-dimensional character: the $q_z$ dependence is
significant and the dominant peak is most intense at $1/4$.  For still larger $\epsilon_x$, coherence begins to be lost.
At $\epsilon_x{=}1.2$\% the IC peak $(0.3,0.3,q_z)$ survives but $\chi^s$ becomes incoherent and diffused over a range
$q$ both in the plane and out of it, with another peak appearing near $(0.15,0.15,0)$.
In short, for $\epsilon_x{>}\epsilon_x^*$, two prominent changes are observed:
incommensurate but nearly ferromagnetic excitations at $(0.15,0.15,0)$ and
commensurate gapped antiferromagnetic spin excitations at $(1/2,1/2,a/2c)$.

\begin{figure}
        \begin{center}
                \includegraphics[width=1.0\columnwidth]{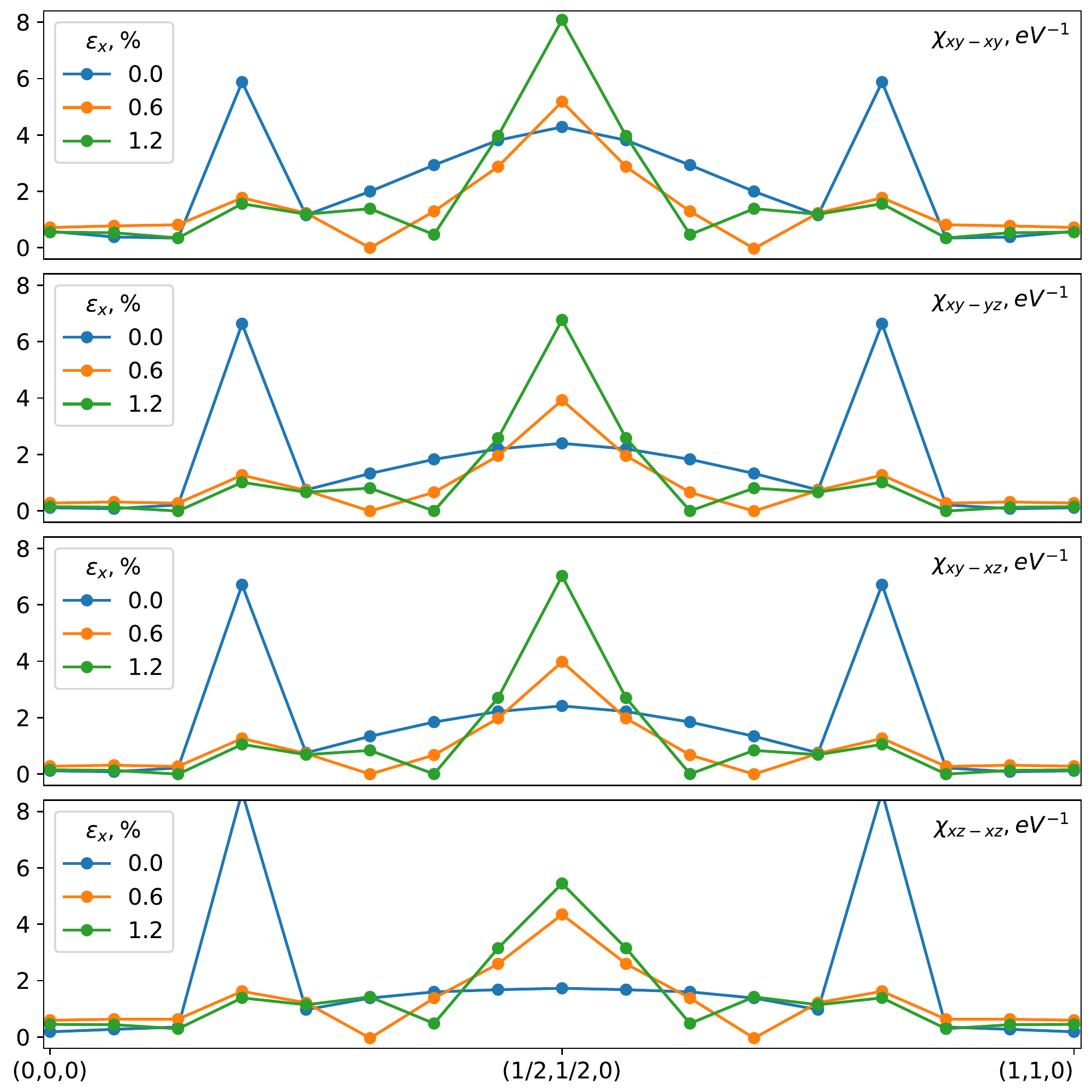}
                \caption{
                                {\bf Charge susceptibilities and commensurability:} Real part of the static charge susceptibility $\chi^{c}(q,\omega=0)$,
                shown along the Cartesian (000) to (110) direction, and for different strains $\epsilon_x$. The unstrained compound shows three-peaked charge fluctuation, with sharp peaks at IC vector (0.2, 0.2, 0) (and by symmetry at (0.8, 0.8, 0)) and a broad peak at (0.5, 0.5, 0). With strain the structure becomes sharply single-peaked at commensurate (0.5, 0.5, 0). The peak at the commensurate vector develops at the cost of the charge fluctuation weights from the IC vectors. The systematic evolution from large wavelength triplet to shorter wavelength singlet fluctuations, under strain, is common to all inter- and intra-orbital charge fluctuations. The strong, often the most dominant, inter-orbital charge fluctuations can be observed in Ru-d$_{xy-xz}$ and Ru-d$_{xy,yz}$ channels.  
                }
                \label{fig:chargefluctuations}
        \end{center}
\end{figure}

\emph{Charge susceptibilities and commensurability:} The evolution of the spin and charge susceptibilities are
instructive to understand the changes in the gap symmetries under strain and their underlying even- or odd-parity
characters. We find that the real part of the charge susceptibilities in the static limit
$\chi^{c}(q,\omega{\rightarrow}0)$, has strong peaks both at the nearly ferromagnetic $\sim$(0.2,0.2,0) vector and also
at more commensurate higher wavelength quasi-anti-ferromagnetic vector (1/2,1/2,0)
(Fig.~\ref{fig:chargefluctuations}).  Raghu et al. discuss a possible route to superconductivity through charge
fluctuations originating from the quasi one-dimensional bands $d_{xz}$ and $d_{yz}$ \cite{kivelson}.  Their analysis
relies on the quasi one-dimensional character of these states.

Our \emph{ab initio} calculation partially supports this picture.  However, we also observe nearly comparable
multi-orbital charge fluctuations, both intra and inter-orbital in nature, in all active bands
(Fig~\ref{fig:chargefluctuations}).  \emph{Inter}-orbital charge fluctuations originating from $d_{xz}$ and $d_{yz}$ and
the two-dimensional $d_{xy}$ are comparable to, or even larger than the intra-orbital contributions.

In the unstrained case, nearly uniform long-wavelength coherent charge fluctuations support a triplet pairing channel
mainly through multi-orbital charge fluctuations.  There is a significant peak in $\chi^q$ at small $q$, near
(0.2,0.2,0).  However, there is another broad peak near (1/2,1/2,0).  Under strain, the latter peak becomes more coherent and
larger, while the former decays.  At the critical strain $\epsilon_x^*$ only the latter peak
at ($\frac{1}{2}$,$\frac{1}{2}$,0) remains. 
Also $\chi^s$ and $\chi^c$ become increasingly coherent close to a vector that favors singlet pairing. This is strikingly different from the unstrained scenario where both spin and charge fluctuations have favourable triplet components as well. For
$\epsilon_x>\epsilon_x^*$, the $\chi^c$ at  ($\frac{1}{2}$,$\frac{1}{2}$,0) becomes large.
Simultaneous shifts in spin fluctuation weight towards more commensurate lower $q$ (larger wavelength) leaves
$\chi^s(q^\mathrm{IC})$ incoherent. This emergent incoherence in spin fluctuations at IC vector is not conducive for pairing.

\emph{Superconducting pairing: nodal character and dimensionality:} 
The superconducting pairing susceptibility $\chi^{p-p}$ is computed by dressing the non-local pairing
  polarization bubble $\chi^{0,p-p}(\textbf{k},i\nu)$ with the pairing vertex $\Gamma^{irr,p-p}$ using the
  Bethe-Salpeter equation in the particle-particle channel.
\begin{eqnarray}
\chi^{p-p} = \chi^{0,p-p}\cdot[\mathbf{1}+\Gamma^{irr,p-p}\cdot\chi^{0,p-p}]^{-1}
\end{eqnarray}
Where the non-local pairing vertex $\Gamma^{irr,p-p}$ in the singlet (s) and triplet (t) channels are
  obtained from the magnetic (spin) and density (charge) particle-hole reducible vertices by
\begin{eqnarray}
\Gamma_{{\alpha_{2},\alpha_{4}\atop \alpha_{1},\alpha_{3}}}^{irr,p-p,s}(\textbf{k},i\nu,\textbf{k}',i\nu') &=& \Gamma_{{\alpha_{2},\alpha_{4}\atop \alpha_{1},\alpha_{3}}}^{f-irr}(i\nu,i\nu')+\frac{1}{2}[\frac{3}{2}\widetilde{\Gamma}^{p-h,(m)}-\frac{1}{2}\widetilde{\Gamma}^{p-h,(d)}]_{{\alpha_{2},\alpha_{3}\atop \alpha_{1},\alpha_{4}}}(i\nu,-i\nu')_{\textbf{k}'-\textbf{k},i\nu'-i\nu}\nonumber\\
&& +\frac{1}{2}[\frac{3}{2}\widetilde{\Gamma}^{p-h,(m)}-\frac{1}{2}\widetilde{\Gamma}^{p-h,(d)}]_{{\alpha_{4},\alpha_{3}\atop \alpha_{1},\alpha_{2}}}(i\nu,i\nu')_{-\textbf{k}'-\textbf{k},-i\nu'-i\nu}
\label{eq:Gamma_pp_nonloc1}
\end{eqnarray}

\begin{eqnarray}
\Gamma_{{\alpha_{2},\alpha_{4}\atop \alpha_{1},\alpha_{3}}}^{irr,p-p,t}(\textbf{k},i\nu,\textbf{k}',i\nu') &=&
 \Gamma_{{\alpha_{2},\alpha_{4}\atop \alpha_{1},\alpha_{3}}}^{f-irr}(i\nu,i\nu')-\frac{1}{2}[\frac{1}{2}\widetilde{\Gamma}^{p-h,(m)}+\frac{1}{2}\widetilde{\Gamma}^{p-h,(d)}]_{{\alpha_{2},\alpha_{3}\atop \alpha_{1},\alpha_{4}}}(i\nu,-i\nu')_{\textbf{k}'-\textbf{k},i\nu'-i\nu}\nonumber\\
 &+&\frac{1}{2}[\frac{1}{2}\widetilde{\Gamma}^{p-h,(m)}+\frac{1}{2}\widetilde{\Gamma}^{p-h,(d)}]_{{\alpha_{4},\alpha_{3}\atop
     \alpha_{1},\alpha_{2}}}(i\nu,i\nu')_{-\textbf{k}'-\textbf{k},-i\nu'-i\nu}
\label{eq:Gamma_pp_nonloc2}
\end{eqnarray}
Finally, $\chi^{p-p}$ can be represented as a function of eigenvalues $\lambda$ and eigenfunctions $\phi^{\lambda}$
of the Hermitian particle-particle pairing matrix (see SM for the detailed derivation).
\begin{eqnarray}
\chi^{p-p}(k,k') = \sum_{\lambda}\frac{1}{1-\lambda}\cdot(\sqrt{\chi^{0,p-p}(k)}\cdot\phi^{\lambda}(k))\cdot(\sqrt{\chi^{0,p-p}(k')}\cdot\phi^{\lambda}(k'))
\end{eqnarray}
The pairing susceptibility diverges when the leading eigenvalue $\lambda$ approaches one. The
  corresponding eigenfunction represents the momentum structure of $\chi^{p-p}$. However, unlike hole doped cuprates or doped single-band Hubbard model~\cite{hyowon},
  the unconventional superconductivity in SRO is multi-orbital in nature with a close packed eigenvalue spectrum, which
  warrants for more detailed investigation of the different eigenfunctions. These eigenfunctions are in different intra- and inter-orbital basis and not embedded into band representation to project the superconducting gap on the Fermi surfaces. To embed these normal phase momentum dependent pairing instabilities in orbital basis to band basis is needed further developments. Once this is done,  we can gap out the Fermi surfaces by diagonalizing a Bogoliubov Hamiltonian built in the band basis~\cite{yin}.

As is apparent from Eqns.~\ref{eq:Gamma_pp_nonloc1},~\ref{eq:Gamma_pp_nonloc2} at what wave vector spin and charge
fluctuations are strong is of central importance to the kind of superconducting pairing symmetry they can form.  If
superconductivity is driven by fluctuations near the ferromagnetic point (0,0,0), the spin part of the Cooper pair is
symmetric and the superconductivity should have triplet symmetry.  If, on the other hand if the fluctuations (spin or
charge) are more proximate to $(1/2,1/2,q_z)$, the symmetry is more likely to be singlet.

\begin{table*}
\begin{tabular}{|p{3cm}|p{1cm}|p{1cm}|p{1cm}|p{1cm}|p{1.6cm}|p{1.0cm}|}
           \hline
             $\Delta_{(\alpha_{1},\alpha_{2}),(\alpha_{3},\alpha_{4})}$  & $\hat{S}$ & $\hat{O}$ &  $\hat{T}$ &
           $\hat{P}$ & pairing functions & irred repsn\\
             \hline
              d$_{(xy,xy),(xy,xy)}$  & -1 &  1 &  1 & 1 & d$_{x^{2}-y^2}$ & B$_{1g}$ \\
              d$_{(xz,xz),(yz,yz)}$  & -1 &  1 &  1 & 1 & S$^{\pm}$      & A$_{1g}$ \\
               d$_{(xz,xz),(yz,yz)}$ & -1 &  1 &  1 & 1 & d$_{x^{2}-y^2}$ & B$_{1g}$ \\
               d$_{(xy,xy),(xy,xy)}$ &  1 &  1 & -1 & 1 & S$^{\pm}$      & A$_{1g}$ \\
               d$_{(xz,xz),(yz,yz)}$ &  1 &  1 & -1 & 1 & S$^{\pm}$      & A$_{1g}$ \\
               d$_{(xz,xz),(yz,yz)}$ &  1 &  1 & -1 & 1 & d$_{x^{2}-y^2}$ & B$_{1g}$ \\
             \hline
\end{tabular}
\label{table1}
\caption{Characterization of different singlet and triplet gap instabilities in terms of D$_{4h}$ irreducible representation. Also shown is how these different gap instabilities transform under spin exchange $\hat{S}$,
orbital exchange $\hat{O}$, time exchange $\hat{T}$ and parity $\hat{P}$ operators.}
\end{table*}

The nature of the candidate superconducting gap structures is strongly debated already in pristine SRO.  Before turning
to the strain dependence, we analyze in detail the gap structures that come out of QS\emph{GW}+DMFT+BSE in both singlet
and triplet channels.  We further analyze how these different gap structures transform under spin exchange $\hat{S}$,
orbital exchange $\hat{O}$, time exchange $\hat{T}$ and parity $\hat{P}$ operators.  This allows us to characterize
their irreducible representations (irreps) in terms of the D$_{4h}$, while noting that under uniaxial strain, the point
group symmetry reduces to D$_{2h}$ and properties of the 4-fold rotational symmetry no longer apply.

\begin{figure}
        \begin{center}
                \includegraphics[width=0.62\columnwidth]{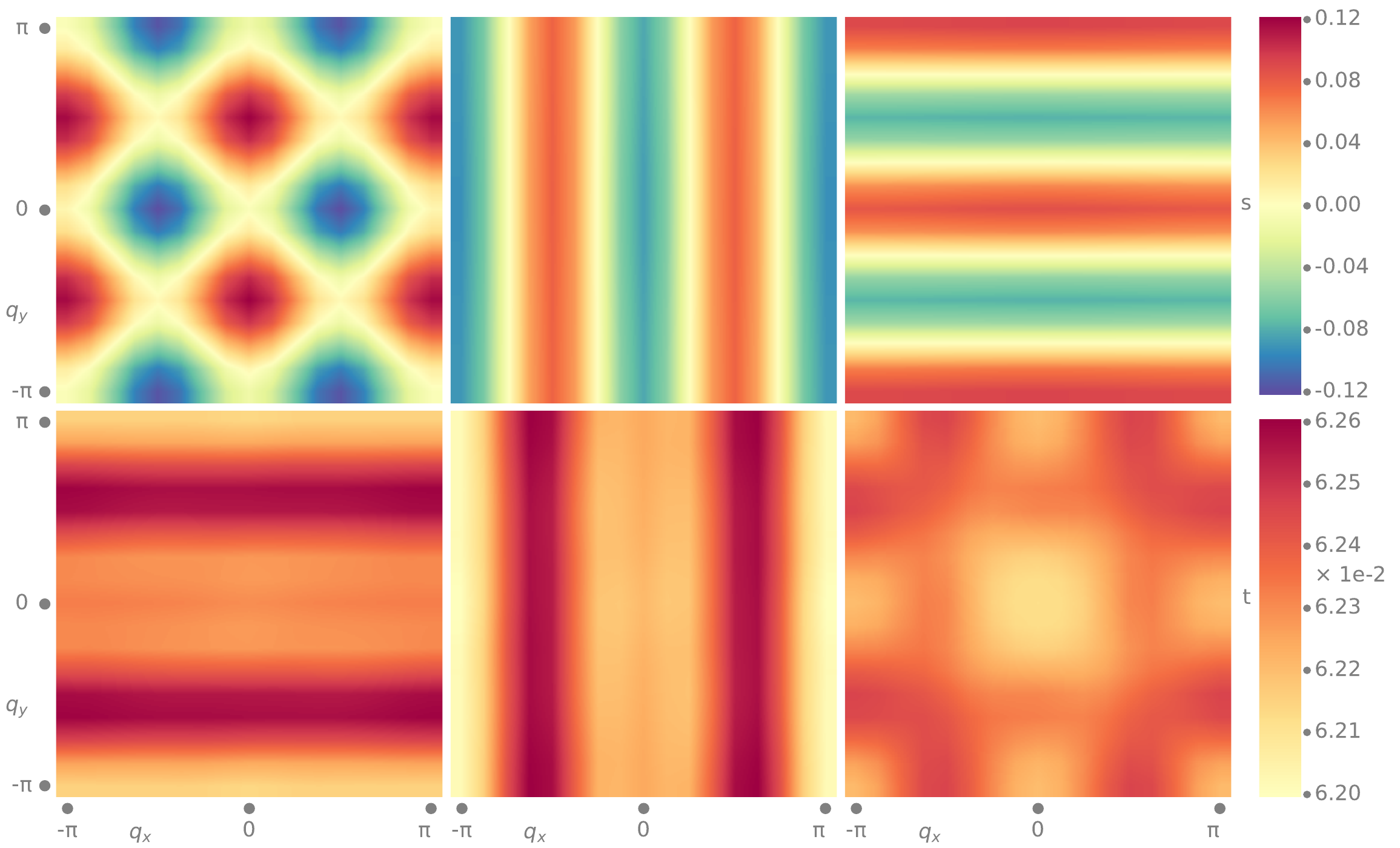}
                \includegraphics[width=0.37\columnwidth]{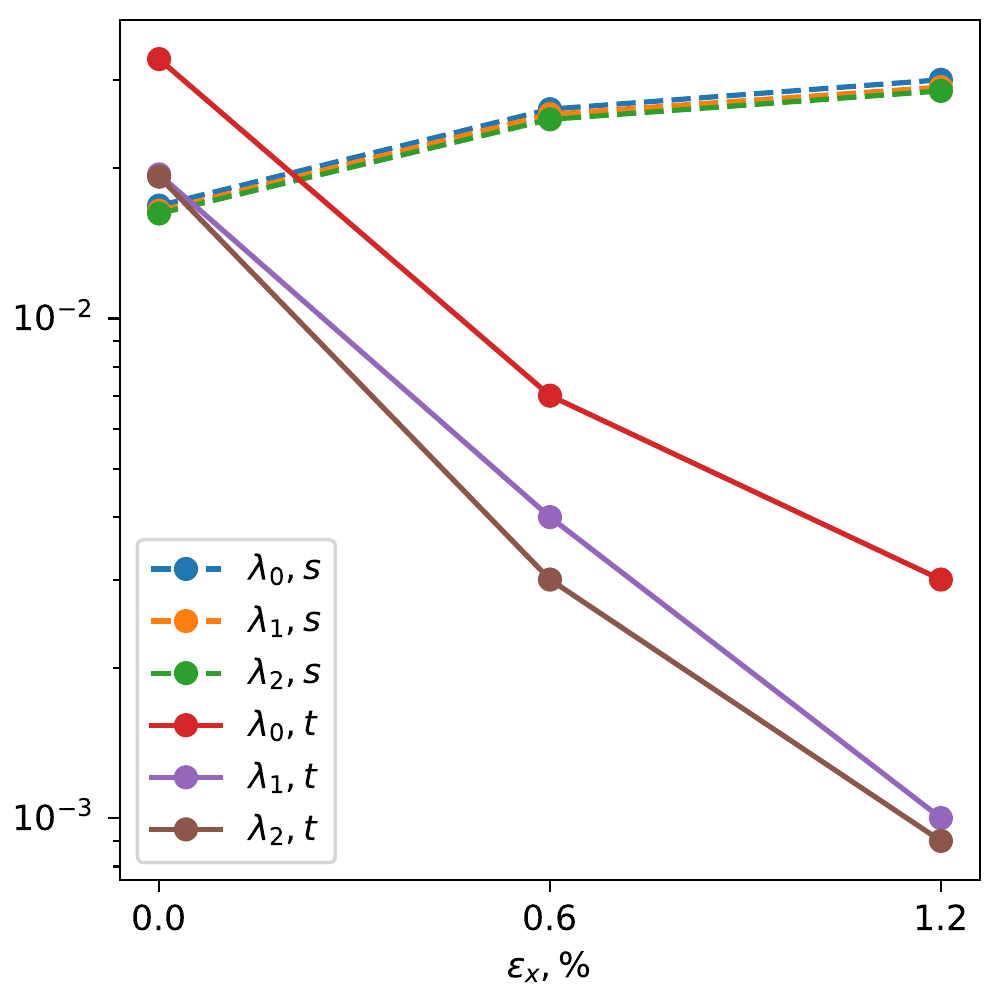}
                \caption{ {\bf Superconducting pairing: nodal character:} (left panel) The superconducting pairing gap symmetries for $\epsilon_x$=0 are shown in the (100)-(010) plane; eigenfunctions corresponding to first three eigenvalues in singlet (s) symmetries are in the top panel and triplets (t) are in the lower panel .  Right panel shows evolution of triplet and singlet eigenvalues under strain. Under strain singlet eigenvalues increase and surpass the triplet eigenvalues.}
                \label{fig:pairing}
        \end{center}
\end{figure}

\emph{Singlet channel:} we find three dominant eigenvalues that contribute to the superconducting gap
  instabilities. The corresponding eigenfunctions are shown in Fig.~\ref{fig:pairing}.  They all change sign; thus the
  gap instabilities have nodes.  For $\epsilon_x{=}0$ we find that all eigenvalues are degenerate within numerical
  precision.  In the Ru-$d_{xy,xy}$ channel, the gap function is a $d$-wave, approximately $\cos k_{x}-\cos k_{y}$, with
  a D$_{4h}$-B$_{1g}$ irreducible representation.  The other two eigenfunctions are respectively $\cos k_{x}$ and $\cos
  k_{y}$ in the $d_{xz,xz}$ and $d_{yz,yz}$ channels.  Hence, if a Cooper pair forms with two quasiparticles belonging
  to two different orbitals then it is possible to get an in-phase extended \emph{s}-wave symmetry with a $d_{xz,yz}$
  gap function $\cos k_{x}+\cos k_{y}$ with irreducible representation A$_{1g}$.  However, it is also possible that
  Cooper pairs form with quasi-particles being out of phase, and hence, leading to a $d_{xz,yz}$ gap function with
  \emph{d}-wave $\cos k_{x}-\cos k_{y}$ symmetry.  Finally, all three eigenvalues increase as temperature drops (see
  SM), suggesting that SRO indeed has instabilities in the spin-singlet channel with nodes that could potentially drive
  Cooper pair formation. This provides a natural explanation for why nodal gap structures are reported from
  several measurements and previous theoretical studies based on model
  Hamiltonians~\cite{linenode18,eremin04,litak04,contreras04,linenode2000,linenode2001,linenode2004}. More importantly
  the new NMR data from Pustogow and Luo et al.~\cite{pustogow} and Ishida et al~\cite{ishida19} have changed the course of discussion significantly in the lines of possible singlet instabilities
  in SRO from the usual triplet routes.  Our study and observation of three singlet eigenvalues present in SRO is
  significant and timely.  However, only recently has pure nodal line gap character been observed experimentally along
  the Cartesian $z$-direction~\cite{hassinger}.  Owing to technical limitations the experiments by Hassinger et
  al.~\cite{hassinger} cannot shed light on the character of the nodal gap structure in the basal plane, but they find
  strong evidence for nodal lines along $q_z$, making a case for the $H_{c2}$ anomaly in SRO.  In a very recent specific
  heat study, under angular variation of magnetic field at very low temperatures, Kittaka et al.~\cite{kittaka}
  established the presence of horizontal line nodes in the gap structure. At this stage we can not shed light onto these
  observations as our analysis of gap instabilities are at present restricted analysis within an orbital basis, as
  perturbations from normal phase.

\emph{Triplet channel:} A very different story emerges.  We find nodeless gap functions as shown in
  Fig.~\ref{fig:pairing}.  As in the singlet case, three eigenvalues are found.  In the $d_{xz,xz}$ and $d_{yz,yz}$
  channels we observe instabilities of the form $\delta_{0}+\cos k_{x}$ and $\delta_{0}+\cos k_{y}$, leading to the
  possibility of an extended nodeless \emph{s}-wave 2$\delta_{0}+\cos k_{x}+\cos k_{y}$ gap structure with A$_{1g}$
  irreducible representation in the $d_{xz,yz}$ basis.  The out of phase coupling between these two quasi-particles will
  lead to B$_{1g}$ nodeless \emph{d}-wave gap structure. We also observe an extended \emph{s}-wave gap function in the
  $d_{xy,xy}$ channel. However, additionally we find a doubly-degenerate set of eigenvalues in an off-diagonal
  $d_{xz,xy}$,$d_{yz,xy}$ channel.  Comparing singlet and triplet channels at fixed temperature, the triplet eigenvalues are nearly two times larger than singlet eigenvalues (see Fig.~\ref{fig:pairing}).  For summary of all gap functions and their irreps see Table 1~\ref{table1}. In short, in the unstrained SRO, triplet channel dominates when it comes to
  Cooper pair formation, however, that does not exclude the possibility of formation of singlet Cooper pairs.

These calculations are performed in the normal phase of the \ce{SRO} ($T{>}1.5$ K), probing all possible
particle-particle instabilities that could precipitate formation of a superconducting gap.  These solutions do not
describe the superconducting state itself, and thus do not include possibility of spontaneously breaking time reversal
symmetry (TRS). However, instabilities that do appear are candidates tailor-made for a complex order parameter
prescribed in earlier theoretical works, notably by Raghu et al.~\cite{kivelson}, Scaffidi et al.~\cite{scaffidi14} and
Mackenzie et al.~\cite{odder}. Such an order parameter can result in a nodeless gap structure which can also lead to TRS
breaking and chiral superconductivity. Also, two-component odd-frequency gap functions can lead to observed Kerr
rotation~\cite{kerr_sro} in \ce{SRO}.  We cannot draw conclusions on these works since at present our \emph{ab initio}
technique limited to discussion of the instabilities towards superconductivity.  However, that the triplet
eigenvalues being larger than the singlet ones in the pristine case strongly suggests triplet pairing dominates.

Remarkably, the roles reverse under strain.  We find that the eigenvalues in the triplet channel decrease rapidly, while
the singlet eigenvalues increase (see Fig.~\ref{fig:pairing}).  At the critical strain $\epsilon_x{=}\epsilon_x^*{=}0.6$\% we observe that the
singlet eigenvalues begin to overtake the triplet eigenvalues, suggesting a strong suppression of the triplet
superconducting instability.  Additionally, we find that the triplet eigenvalues are weakly dependent on temperature at
this critical $\epsilon_x^*$ while the singlet eigenvalues start to diverge with lowering temperatures. This implies
that the superconducting state switches from being dominated by triplet pairing at $\epsilon_x{=}0$ to singlet pairing
at $\epsilon_x{=}\epsilon_x^*$.  This is fully consistent with the susceptibility calculations, which suggest that under
strain both spin and charge fluctuations becomes more intense close to singlet vectors, and far from triplet $q=0$.

\emph{Spin-charge co-operation in singlet channel under strain:} To address the issue conclusively we compute the eigenvalues with and without the charge vertex functions. With both spin and charge vertex functions (as the calculations should be performed), we observe that for $\epsilon_{x}$=0 triplet eigenvalues are larger than the singlet eigenvalues. On suppressing the charge vertex, we find that for $\epsilon_{x}$=0, it does not have any effect on the result; triplet eigenvalues remain larger than the singlet eigenvalues. However, for finite strain, $\epsilon_{x} > 0$, once the charge vertex is suppressed the triplet eigenvalues become larger than the singlet eigenvalues, which is in contrary to the observation including both spin and charge vertex functions. This is a strong indication that both spin and charge vertex functions co-operate under strain in the singlet channel. In Table 2~\ref{table2} we show the eigenvalues with and without the charge vertex to make a case for this argument. 

\begin{table}
	\begin{center}
		\footnotesize
		\begin{tabular}{ccccccccccc}
			\hline
			variants & $\lambda_{s}^{f}$ & $\lambda_{t}^{f}$ & $\lambda_{s}^{-c}$ & $\lambda_{t}^{-c}$ \\
			\hline
		
			 $\epsilon_{x}$=0  & 0.006 & 0.011 & 0.005 & 0.014  \\
			
			$\epsilon_{x}> 0$ & 0.011 & 0.007 & 0.014 & 0.032 \\
			
								\end{tabular}
	\end{center}
	\label{table2}
	\caption{The leading singlet and triplet eigenvalues computed using both spin and charge vertex functions $\lambda_{s}^{f}$, $\lambda_{t}^{f}$, and by suppressing the charge vertex functions  $\lambda_{s}^{-c}$, $\lambda_{t}^{-c}$.}
\end{table}

\emph{Conclusion:} multiple singlet and triplet superconducting instabilities are observed in SRO.  A purely spin
triplet superconductivity needs sufficient coherent and low energy spin fluctuation glue near ferromagnetic vector
$q$=0. However, our results show that the dominant spin fluctuations are at $(0.3,0.3,q_z)$ which is closer to the
singlet-pairing vector, combined with the smaller peak at the quasi-ferromagnetic `triplet' vector. Multi-orbital charge
correlations also play a central role in \ce{Sr2RuO4}: they provide additional glue both at low-q and $(0.5,0.5,q_z)$
through strong intra- and inter-orbital fluctuations. Together, they lead to multiple triplet and singlet Cooper pair
instabilities, although triplet eigenvalues are larger than the singlet eigenvalues.

When strain is applied the dominant character of Cooper pair instability changes.  $\chi^{s}(q{=}0.3,0.3,q_z)$ becomes
more coherent up to a critical strain $\epsilon_x^*$.  Simultaneously the spectral weight under the low-q charge peak in
gets fully transferred to a more coherent quasi-anti-ferromagnetic vector $(0.5,0.5,q_z)$.  Together they suggest, spin
and charge co-operate to sustain an even parity pairing channel which maximize $T_{c}$ at $\epsilon_x^*$. For
$\epsilon_x{>}\epsilon_x^*$, the spin fluctuation weight drifts toward larger wavelength, more uniform
quasi-ferromagnetic vectors and charge fluctuates more strongly at the quasi-anti-ferromagnetic vector. This emergent
spin incoherence and spin-charge separation, split by quasi-ferromagnetic spin fluctuation peak and
quasi-anti-ferromagnetic charge fluctuation peak, is not conducive for sustaining the even-parity superconductivity and
hence lowers and suppresses $T_{c}$. Our observations suggest that the pathway to maximize superconductivity in
\ce{Sr2RuO4} would be to cause spin and charge fluctuations to act in symphony in an even parity channel.


\section*{Acknowledgments}
Authors acknowledge F. Baumberger for sharing with us the raw ARPES data for Fermi surfaces. SA acknowledges discussions with James Annett, Martin Gradhand and Astrid Romer. This work was supported by the Simons Many-Electron Collaboration,
and EPSRC (grants EP/M011631/1 and EP/M011038/1).
For computational resources, we were supported by the ARCHER UK National Supercomputing Service, UK Materials and Molecular Modelling Hub for computational resources, (EPSRC grant EP/P020194/1) and computing resources provided STFC Scientific Computing Department's SCARF cluster and PRACE supercomputing facility. 

\section*{Method}

We use a recently developed quasi-particle self consistent \emph{GW} + dynamical mean field theory
(QS\emph{GW}+DMFT)~\cite{prx18,prbr17}, as implemented in the all-electron Questaal package \cite{questaal_web}. Paramagnetic DMFT is
combined with nonmagnetic QS\emph{GW} via local projectors of the Ru 4$d$ states on the Ru augmentation spheres to form
the correlated subspace.  We carried out the QS\emph{GW} calculations in the tetragonal and strained phases of
\ce{Sr2RuO4} with space group 139/I4mmm.  DMFT provides a non-perturbative treatment of the local spin and charge
fluctuations. We use an exact hybridization expansion solver, namely the continuous time Monte Carlo (CTQMC)
\cite{Haule_long_paper_CTQMC}, to solve the Anderson impurity problem.

The one-body part of QS\emph{GW} is performed on a $16 \times 16 \times 16$ k-mesh and charge has
been converged up to $10^{-6}$ accuracy, while the (relatively smooth) many-body static self-energy $\Sigma^0(\mathbf{k})$
is constructed on a $8 \times 8 \times 8$ k-mesh from the dynamical $GW$ $\Sigma(\mathbf{k},\omega)$.
$\Sigma^0(\mathbf{k})$ is iterated until convergence (RMS change in $\Sigma^0{<}10^{-5}$\,Ry).
$U$=4.5\,eV and $J$=1.0\,eV \cite{ujkotliar} were used as correlation parameters for DMFT. The DMFT for the dynamical self energy is iterated, and converges in $\approx 20$ iterations.
Calculations for the single particle response functions are performed with $10^9$ QMC steps per core and the statistics
is averaged over 64 cores. The two particle Green's functions are sampled over a larger number of
cores (40000-50000) to improve the statistical error bars. We sample the local two-particle Green's functions
with CTQMC for all the correlated orbitals and compute the local polarization
bubble to solve the inverse Bethe-Salpeter equation (BSE) for the local
irreducible vertex. 
Finally, we compute the non-local polarization bubble $G(\mathbf{k},\omega)G(\mathbf{k}{-}\mathbf{Q},\omega{-}\Omega)$
and combined with the local irreducible vertex \cite{hywon_vertex} we obtain the full non-local spin and charge
susceptibilities $\chi^{s,c}(\mathbf{Q},\Omega)$. The susceptibilities are computed on a $16 \times 16 \times 16$
$\mathbf{Q}$-mesh. BSE equations in the particle-particle pairing channels are solved on the same k-mesh to extract the
susceptibilities and the Eliashberg eigenvalue equations are solved to extract the eigenvalue spectrum and corresponding
pairing symmetries~\cite{hyowon}.

\section*{Author contributions}
SA has conceived and designed the research. SA, MvS have carried out
the calculations. SA, DP, HP, MvS have contributed codes. DP, MvS have prepared the figures. All authors have contributed to the writing of the paper and the analysis of the data.

\section*{Additional information}
Supplementary material is available.

\section*{Competing financial interests}
The authors declare no competing financial interests.

\section*{Correspondence}
All correspondence, code and data requests should be made to SA. 

%
%

\clearpage



\section{Supplemental Material}

In the supplementary material for our main paper titled `Evening out the spin and charge parity to increase T$_c$ in unconventional superconductors', we discuss structural inputs for our calculations, the non-trivial properties of SRO density of states (DOS), the relative orientations of QS\emph{GW} energy levels, method for computation of superconducting gap structure and the importance of non-local electronic correlations in SRO. These discussions are beyond the limited scope of the main text, while help in understanding the technical details of our implementations and their relevance in understanding the physics discussed in the main paper. 

\ce{Sr2RuO4} has a BCT lattice structure.  We took the lattice parameters and internal positions
  from~\cite{sr214structure}. The lattice constant in the basal plane is $a$=3.86448\,\AA. The uniaxial strain was performed at constant volume.

\begin{figure}
        \begin{center}
                 \includegraphics[width=0.48\columnwidth]{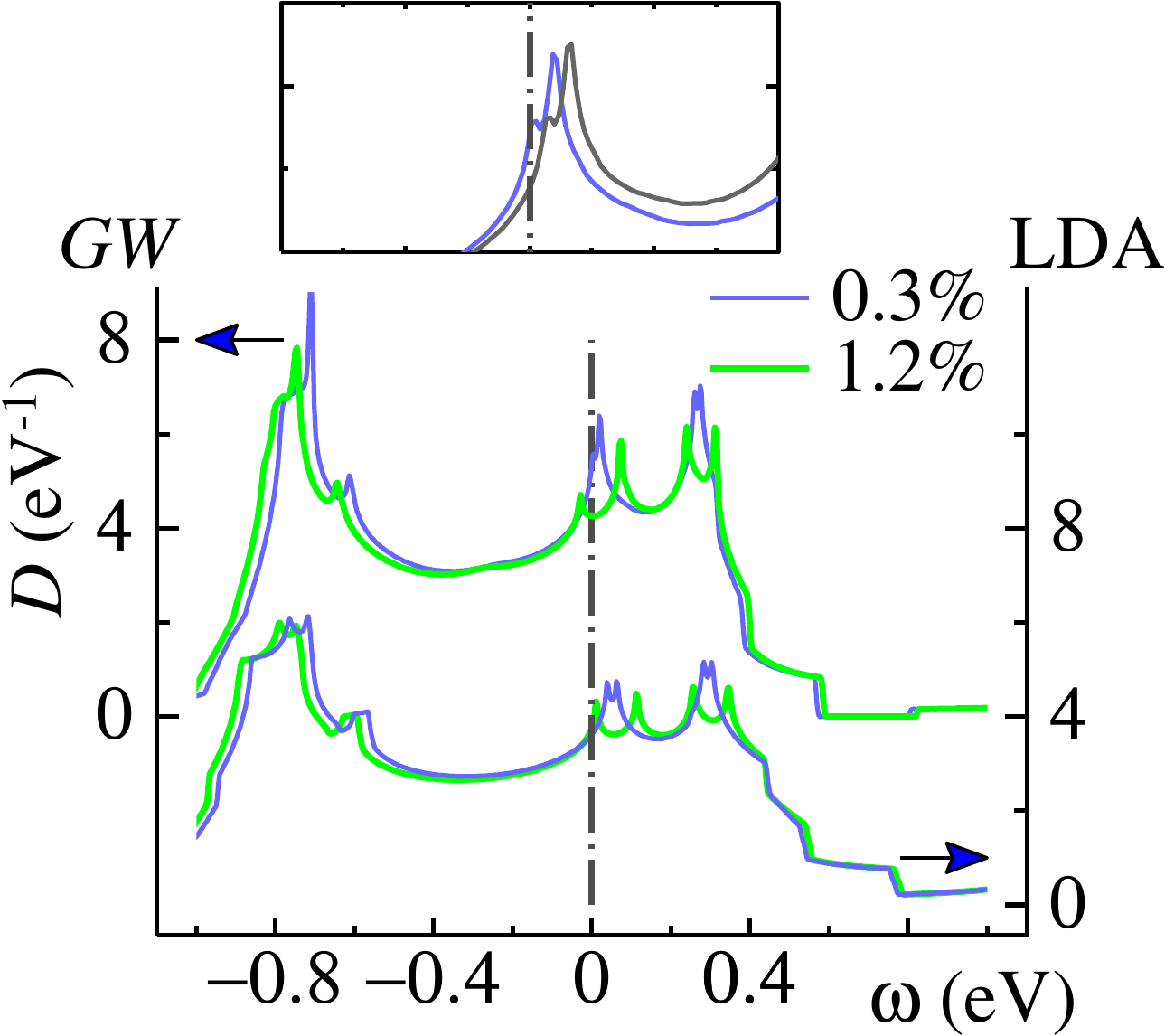}
                 \includegraphics[width=0.48\columnwidth]{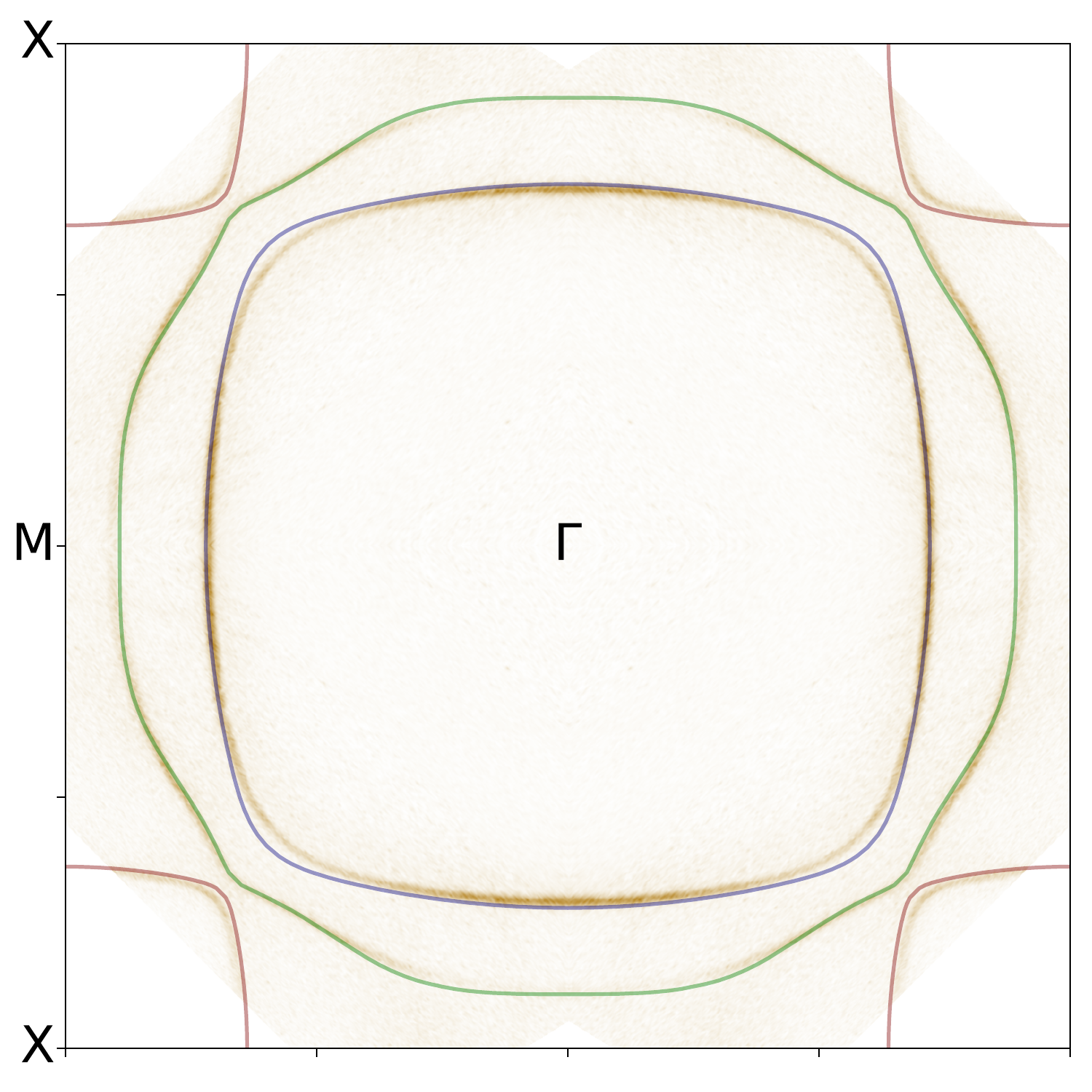}
                        \caption{ {\bf Density of States:} Top pair of curves show density of states (DOS) from
                        QS\emph{GW}, for strains $\epsilon_x$ = 0.3\% (blue) and 1.2\% (green).  A Van-Hove singularity
                        appears slightly above the Fermi level when $\epsilon_x$=0.  When $\epsilon_x{\ne}0$ the
                        singularity splits into two peaks, with the smaller peak crossing $E_{F}$ at
                        $\epsilon_x{=}0.6\%$.  An intriguing possibility is to find conditions that cause the larger
                        peak to cross $E_{F}$.  Bottom pair is the corresponding DOS in the LDA.  The Van-Hove
                        singularity and its splitting are also seen, but the peaks split more symmetrically, and evolve
                        more slowly with $\epsilon_x$.  On average, the LDA DOS is 25\% smaller than the QS\emph{GW}
                        DOS, which is a consequence of the LDA's tendency to overestimate \emph{d} bandwidths.  Inset
                        shows the QS\emph{GW} DOS at $\epsilon_x{=}0.3\%$ on finer energy scale.  The grey line shows
                        the DOS with spin-orbit coupling removed. To the right, the high-resolution ARPES data~\cite{tamai2018} for the Fermi surfaces are shown (figure reproduced with due permission) in the background of our QS\emph{GW} theoretical data. Weak discrepancies can be observed for the quasi-one-dimensional Fermi sheets.}
                 \label{dos}             

        \end{center}
\end{figure}

\emph{Local density of states (DOS):}
The DOS obtained from QS\emph{GW} shows the presence of a sharp resonance slightly above the Fermi
level in the unstrained case (Fig.~\ref{dos}).  Strain splits the single peak, the splitting increasing linearly with
$\epsilon_x$.  The smaller, lower peak crosses $E_{F}$ at $\epsilon_x$=0.6\%, which precisely coincides with the critical
strain $\epsilon^*_x$ where $T_C=T^\mathrm{max}_C$.

\begin{figure}
        \begin{center}
                \includegraphics[width=1.0\columnwidth]{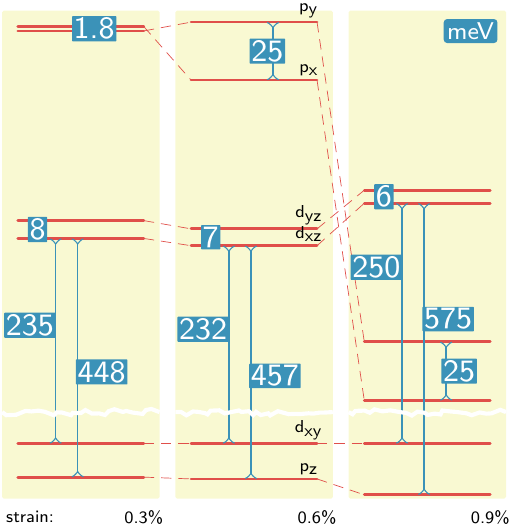}
                \caption{
                        {\bf Energy levels:} Relative orientations of different active orbitals in $\ce{Sr2RuO4}$ from QS\emph{GW}. In strained $\ce{Sr2RuO4}$, we find typical splitting of 6-8 meV between Ru-d$_{xz}$ and d$_{yz}$ orbitals and about 25 meV between O-p$_{x}$ and O-p$_{y}$. This energy splitting is larger than the spin exchange scale in $\ce{Sr2RuO4}$.
                }
                \label{levels}
        \end{center}
\end{figure}

\emph{Energy levels:} Strain lifts the tetragonal symmetry, splitting the degeneracy of the Ru
(4d$_{xz}$,\,4d$_{yz}$) and O (p$_{x}$,\,p$_{y}$) pairs. We find that the states split beyond the typical exchange scale
of $\sim$4\,meV~(Fig. \ref{levels}) when $\epsilon_x$=$0.6\%$.  The split between Ru
4d$_{xz}$,\,4d$_{yz}$ is extremely small, and might fall below the resolution of several measurements trying to probe directly the consequences of lifting of the native teragonal symmetry under strain. However, O p$_{x}$,\,p$_{y}$ states split by nearly 20-25 meV. 

\begin{figure}
  \begin{center}
        \includegraphics[width=1.0\columnwidth]{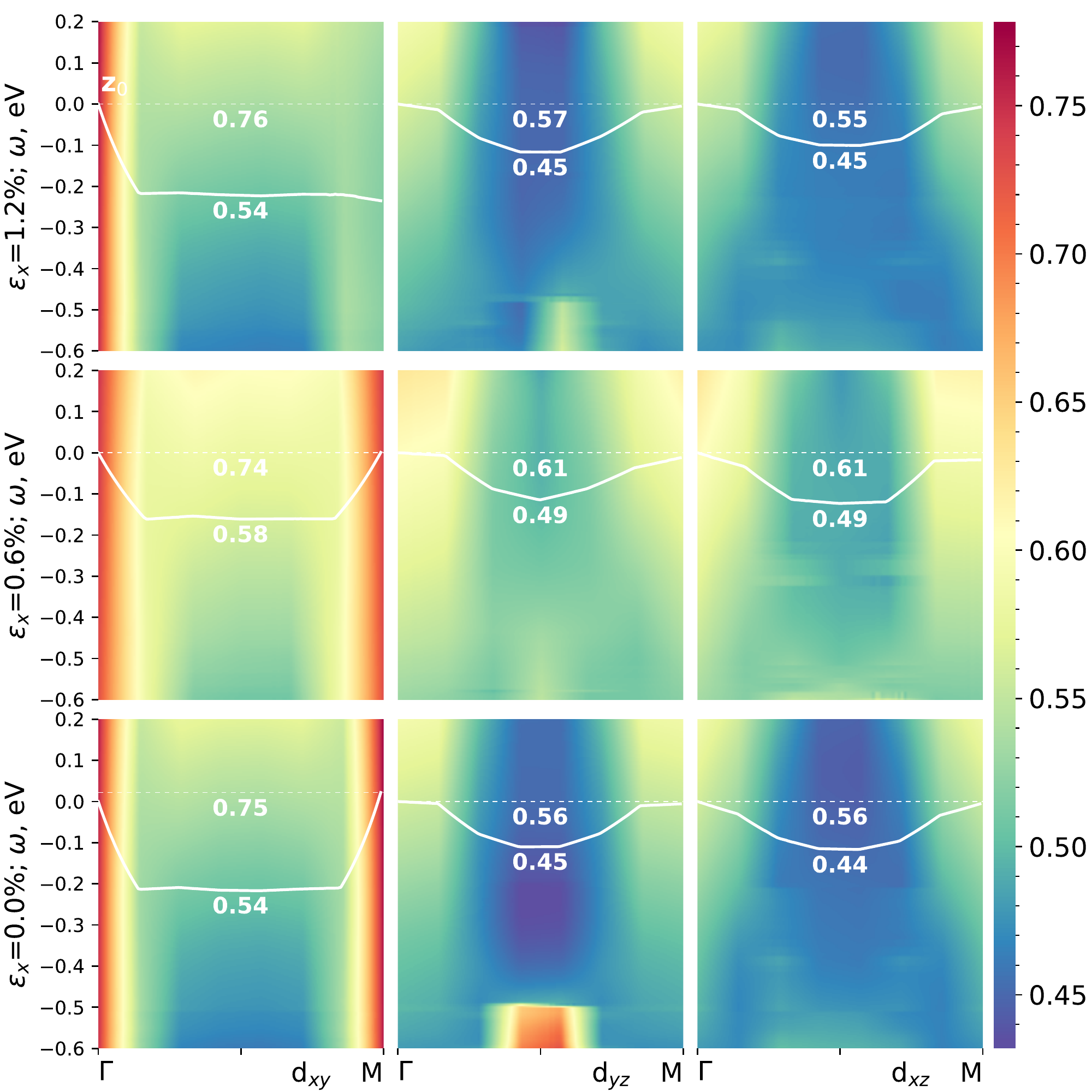}
        \caption{{\bf Non-local correlations:} Quasi-particle renormalization factor, $Z_{k}(\omega)$, computed in QS\emph{GW}, as a heat map in the $k{-}\omega$ plane. White lines detail $Z$ at the Fermi energy ($\omega=0$) along the $\Gamma$-M line.  The three bands present at
                $E_{F}$ are all of Ru t$_{2g}$ character, respectively d$_{xy}$, d$_{yz}$, d$_{xz}$-like on this line (from left to right).  \emph{Z} is a measure of how strongly nonlocality in space and time
                renormalize and smear out the energy bands: at $Z{=}1$ electrons are perfectly coherent and act like
                independent particles; when $Z{=}0$ all coherence is lost.  Its $k$-dependence is unusually strong.
                To put it in perspective, in iso-structural \ce{La2CuO4}, $Z$ varies by $\sim$20\%~\cite{prx18} for the Cu-d$_{x^2-y^2}$ band.
                Lower panels show $Z_k$ evolves in a non-trivial, orbital-dependent fashion with strain.
                Spin fluctuations missing in QS\emph{GW} further reduce $Z$.
                }
                \label{nonlocal}
 \end{center}
\end{figure}

\emph{Non-local Coulomb correlations:} As Coulomb interactions are long-range, they are treated by QS\emph{GW} perturbatively in a self-consistent manner through
a dynamic and momentum dependent self-energy, $\Sigma(\mathbf{k},\omega)$.  We analyse $\Sigma(\mathbf{k},\omega)$
computed within QS\emph{GW} to extract the momentum-dependent quasi-particle renormalization factor
$Z_{k}=(1-{\partial\Sigma(\mathbf{k},\omega)}/{\partial\omega})^{-1}$.  Fig.~\ref{nonlocal} shows how the non-local
$\Sigma(\mathbf{k},\omega)$ renormalizes bands differently on the $\Gamma$-M line.  The variation is
orbital-dependent, fluctuating by ${\sim}40\%$ in a nontrivial manner with $\mathbf{k}$, $\omega$, and strain.
This has a significant effect on the $d$ bandwidth, and is one important reason why QS\emph{GW} bands are 
significantly narrower than their LDA counterparts.  Bandwidths are nevertheless much larger than ARPES measurements, and
they get further renormalized by strong spin fluctuations included in DMFT.

\begin{figure}[!htbp]
\begin{center}
\includegraphics[scale=0.26]{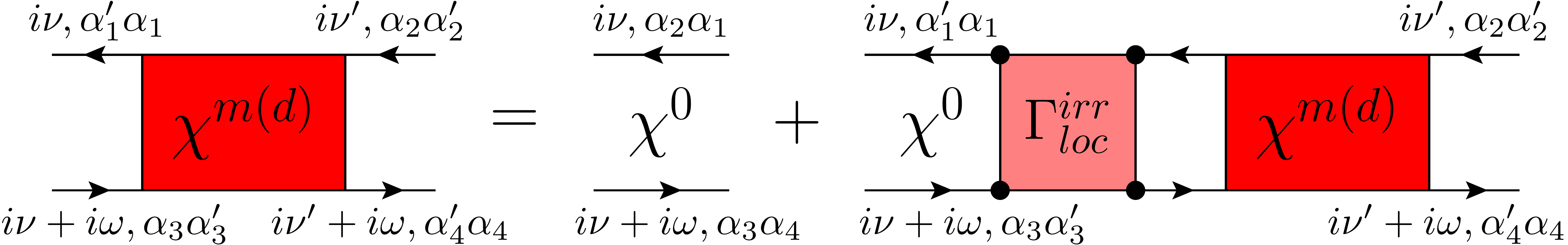}
\includegraphics[scale=0.66]{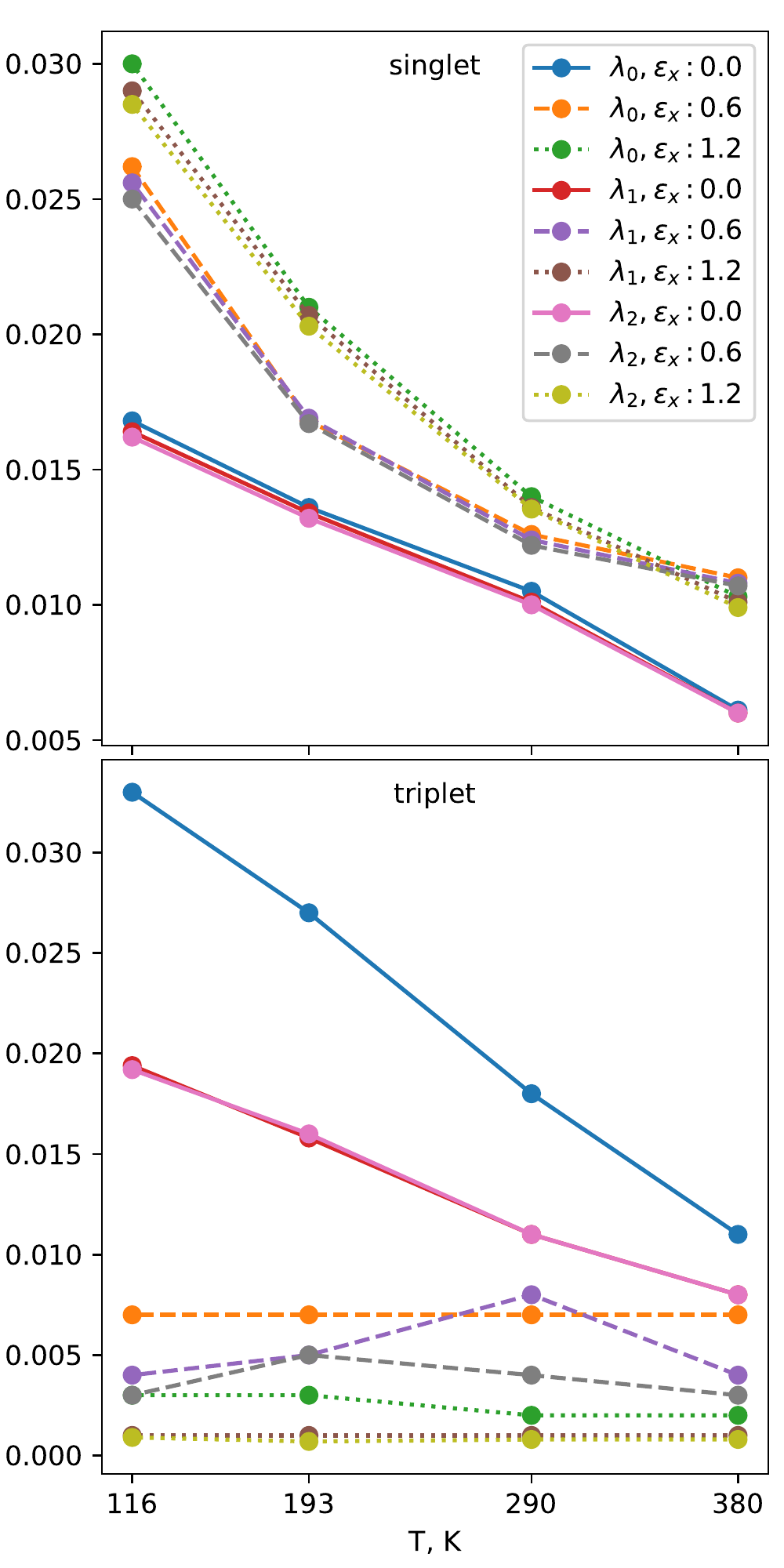}
\caption{The Feynman diagram for the Bethe-Salpeter equation in the spin (charge)
channel. The non-local susceptibility is obtained by replacing the
local propagator by the non-local propagator.\label{fig:BSE-diagram}. In lower panel first ($\lambda_{0}$), second ($\lambda_{1}$) and third ($\lambda_{2}$) eigenvalues with and without strain as function of temperature in singlet and triplet channels are shown. }
\end{center}
\end{figure}

\emph{Magnetic, charge and Superconducting susceptibilities:}
Here we show the Feynman diagram representation of the Bethe-Salpeter equation in the particle-hole (p-h) channel~\ref{fig:BSE-diagram}. 

In order to extract $\Gamma_{loc}^{irr}$, we employ the
Bethe-Salpeter equation which relates
the local two-particle Green's function ($\chi_{loc}$) sampled by
CTQMC, with both the local polarization function ($\chi_{loc}^{0}$)
and $\Gamma_{loc}^{irr}$. 
\begin{equation}
\Gamma_{loc{\alpha_{1},\alpha_{2}\atop \alpha_{3},\alpha_{4}}}^{irr,m(d)}(i\nu,i\nu^{\prime})_{i\omega}=[(\chi_{loc}^{0})_{i\omega}^{-1}-\chi_{loc}^{m(d)-1}]_{{\alpha_{1},\alpha_{2}\atop \alpha_{3},\alpha_{4}}}(i\nu,i\nu^{\prime})_{i\omega}.
\end{equation}
The non-local polarization bubble in the p-h channel is computed from
\begin{equation}
\chi_{{\alpha_{1}\sigma_{1},\alpha_{2}\sigma_{2}\atop \alpha_{3}\sigma_{3},\alpha_{4}\sigma_{4}}}^{0}(i\nu,i\nu')_{\textbf{q},i\omega} = -\frac{T}{N_{k}}\sum_{\textbf{k}}G_{\alpha_{2}\alpha_{1},\sigma_{1}}(\textbf{k},i\nu)\cdot G_{\alpha_{3}\alpha_{4},\sigma_{3}}(\textbf{k}+\textbf{q},i\nu+i\omega)\cdot\mbox{\ensuremath{\delta}}_{i\nu,i\nu'}\cdot\delta_{\sigma_{1}\sigma_{2}}\cdot\delta_{\sigma_{3}\sigma_{4}}\label{eq:nonlocal_bub}
\end{equation}
With these two quantities we can get the full dynamic and non-local spin and charge susceptibilities. 

Further, we show the Feynman diagram representation of the p-p vertex and it's decomposition using the p-h vertex function computed in the magnetic and density channels.

\begin{figure}[!htbp]
\begin{center}
\includegraphics[scale=0.22]{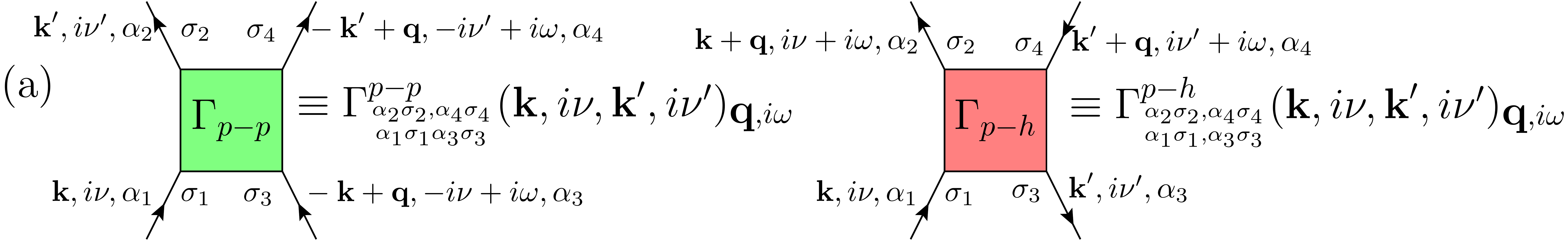}
\includegraphics[scale=0.25]{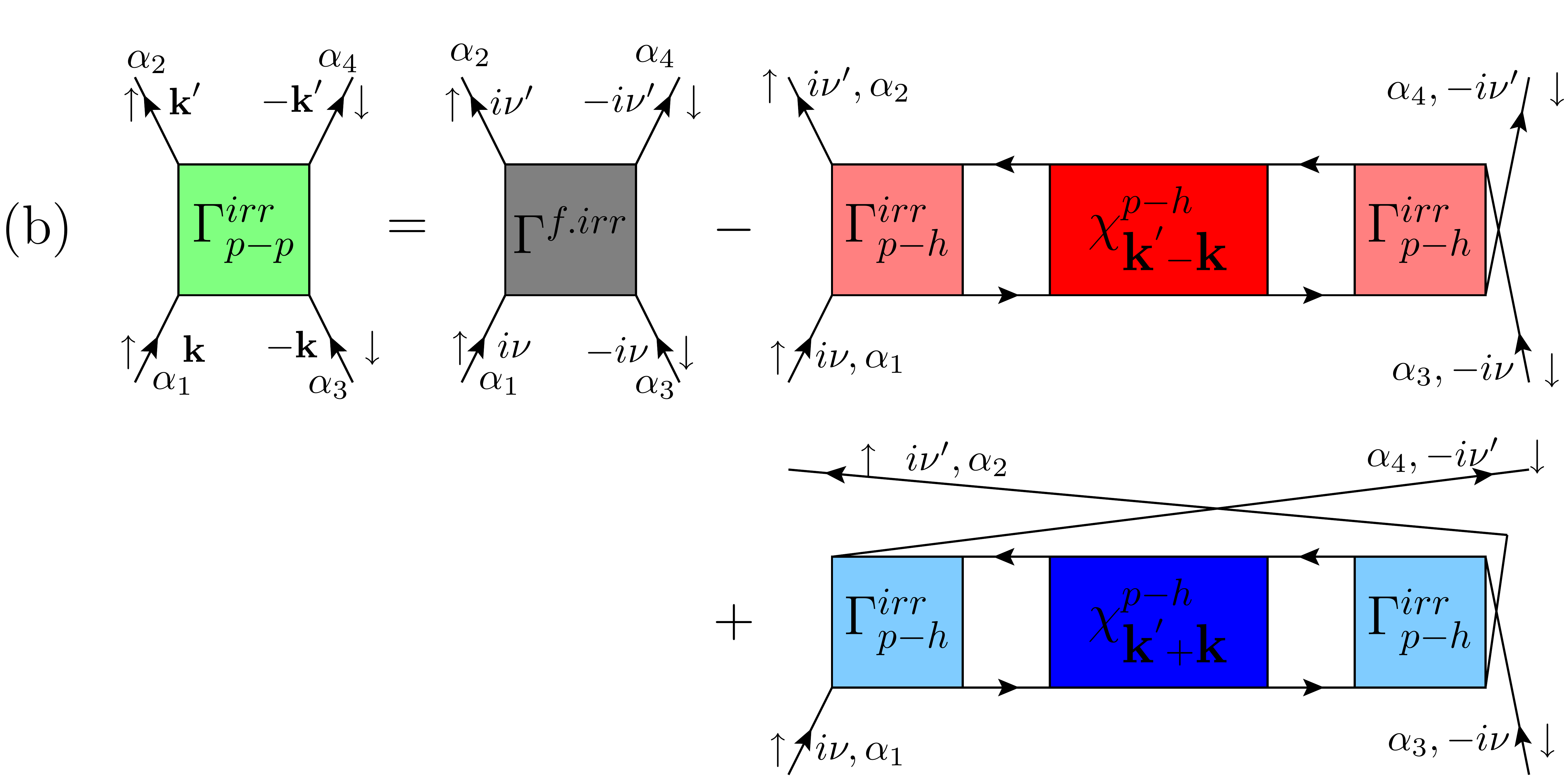}
\caption{(a) The spin, orbital, momentum, and frequency labels of vertex functions
in the particle-particle ($p-p$) channel and the particle-hole ($p-h$)
channel. (b) The decomposition of the irreducible vertex function
in the particle-particle channel $\Gamma^{irr,p-p}$. \label{fig:Gamma_pp} }
\end{center}
\end{figure}
\end{document}